\documentclass[traditabstract]{aa}

\usepackage{graphicx}
\usepackage[pdftex]{hyperref}
\usepackage{txfonts}
\usepackage{natbib}

\bibpunct{(}{)}{;}{a}{}{,}
\begin{document}

   \title{The Galactic dust-to-metals ratio and
          metallicity using gamma-ray bursts}

   \subtitle{}

   \titlerunning{The Galactic dust-to-metals ratio using GRBs}

   \author{D.~Watson\inst{1}
          }

   \authorrunning{D.~Watson}

   \institute{Dark Cosmology Centre, Niels Bohr Institute, University of
              Copenhagen, Juliane Maries Vej 30, DK-2100 Copenhagen \O, Denmark\\
              \email{darach@dark-cosmology.dk}
             }

   \date{Received XXX; accepted XXX}

  \abstract
  {The metallicity and dust-to-metals ratio of the Galaxy are
   fundamental parameters in understanding the interstellar medium (ISM). 
   Currently, there is still some uncertainty surrounding these parameters. 
   In this paper, the dust-to-metals ratio in the Galaxy is determined using
   the photoelectric absorption of the soft X-ray afterglows of a large
   sample of several hundred gamma-ray bursts (GRBs) to determine the metal
   column density in combination with Galactic dust maps to determine the
   line-of-sight dust extinction through the Galaxy in the direction of the
   GRB.  GRB afterglows often have large extragalactic
   soft X-ray absorptions and therefore the GRB sample's upper-bound will
   define the Galactic dust-to-metals relation.  Using a two-dimensional
   two-sample Kolmogorov-Smirnoff test, we determine this upper-bound and so
   derive the dust-to-metals ratio of the Galaxy.  We find $A_V =
   2.2^{+0.3}_{-0.4}\times10^{21}$\,cm$^{-2}$ $N_{\rm H}$ assuming solar,
   Anders \& Grevesse (1989), metallicity.  This result is consistent with
   previous findings using bright X-ray sources in the Galaxy.  Using the
   same technique but substituting the \ion{H}{i} maps from the
   Leiden-Argentine-Bonn survey for the dust maps, allows us to place a
   limit on the metallicity in the Galaxy.  We find a metallicity consistent
   with the Anders \& Grevesse (1989) solar values often used in X-ray
   fitting.  Based on this and previous studies, we suggest that the
   metallicity of a typical ISM sightline through the Galaxy is
   $\sim0.25$\,dex higher than the current best estimate of the solar
   metallicity.  We further show that the dust-to-gas ratio seems to be
   correlated with the total gas column density, and that this may be due to
   the metallicity gradient observed toward the Galactic centre.  Based on
   the non-constant nature of the dust-to-gas ratio, we propose that the
   dust column density, at $N_{\rm H} = 2.2\times10^{21}$\,cm$^{-2}$\,$A_V$,
   represents a better proxy for the soft X-ray absorption column density
   than \ion{H}{i} maps.}

   \keywords{ Gamma-ray burst: general -- ISM: abundances -- dust, extinction
              -- Galaxy: abundances -- X-rays: ISM
               }

   \maketitle

\section{Introduction}
 
The relationship between gas, metals and dust that defines the
interstellar medium (ISM), plays a central role in the properties of
star-formation, and in the appearance, evolution and ultimate fate of
galaxies.  However the basic quantitative relationship between gas, metals
and dust is still not well-defined even for our own galaxy.  Many studies
have been made over the years examining the dust-to-gas ratio in the Galaxy
and more recently at cosmological distances.  Two methods dominate these
analyses: a) comparing hydrogen absorption to dust extinction or reddening,
using the Ly$\alpha$ line in the UV (and sometimes H$_2$ UV lines) to get
the total gas column and pairs of stars to obtain the total extinction, and
b) soft X-ray photoelectric absorption that measures the total metal column
density in the foreground of bright X-ray sources, and converting this to a
gas column assuming a metallicity, and comparing this to a dust extinction
obtained from methods like the Balmer decrement, the deviation in the
optical/infrared from a blackbody spectrum or even measuring the dust column
using the halo made by small angle scattering of X-rays off the dust.  The
first method provides a genuine gas-to-dust ratio, in the sense that it
measures the bulk of the gas directly.  However, it suffers from the
deficiency that it is insensitive to ionised gas and unless the molecular
lines are measured, also to H$_2$, where the fraction of hydrogen in
H$_2$ may be $\sim 0.5$ for $A_V\gtrsim0.5$ \citep{2009ApJS..180..125R}. 
Furthermore it only works along relatively low extinction lines of sight
($A_V\lesssim2-3$) since it requires spectroscopy in the UV
where the extinction is far higher and stars are often too faint to observe
in Ly$\alpha$ or H$_2$.  The X-ray absorption method works to very high
extinctions ($A_V\sim30$ or more) and measures essentially all metals
whether they are ionised or even in the solid phase, providing a census of
the total column density in metals.  However, it is a measurement of the
metal column, not the gas column since the X-ray absorption is almost
insensitive to hydrogen absorption and depends only weakly on helium. 
Therefore the X-ray absorption requires a metallicity conversion to move
from a dust-to-metals to a dust-to-gas ratio.

It is perhaps worth noting that most studies to date in the Galaxy have
either focussed on relatively nearby sets of objects or provided a small
number (between 3 and about 20 for the X-ray studies) of lines of sight to
locations within a few kpc of the Sun. Most studies in the Galaxy have
therefore not probed the ISM of the Milky Way in either a complete or
unbiased way.  These studies have consistently found a dust-to-gas ratio at
a level of $N_H/A_V \sim 2\times10^{21}$\,cm$^{-2}$\,mag$^{-1}$
\citep{1978ApJ...224..132B,1981MNRAS.196..469W,1994ApJ...427..274D,1996Ap&SS.236..285R,1973A&A....26..257R,1975ApJ...198...95G,1975ApJ...198..103R,1995A&A...293..889P,2003A&A...408..581V,2009MNRAS.400.2050G,2009ApJS..180..125R}.
The statistical errors quoted for some studies have been as small as a few
$10^{19}$\,cm$^{-2}$\,mag$^{-1}$, however the variation from study to study
is closer to a few parts in $10^{20}$\,cm$^{-2}$\,mag$^{-1}$ 
This discrepancy may be related to an underestimate of the uncertainties or to an
intrinsic scatter in the relation.

Gamma-ray bursts (GRBs), while found at cosmological distances, are
extremely bright.  In this paper we use the large homogeneous sample of
\emph{Swift} GRB X-ray afterglows to determine upper limits to the metal
column densities to several hundred lines of sight through the Galaxy.  We
compare these metal column densities to all-sky hydrogen and dust surveys to
obtain a new dust-to-metals ratio and metallicity value for the Galaxy.  The
GRB afterglows are subject to absorption by their hosts, with a
minor contribution from intervening objects, so we use a 2D 2-sided
Kolmogorov-Smirnov (KS) test to overcome this limitation to define the
Galactic lower envelope to their absorbing column densities.  Since it is
X-ray selected, the sample is not afflicted by observational bias as UV
studies are.  However, its greatest benefit over previous samples is that
this sample passes lines of sight at random through the Galaxy, and passes
through the entire Galaxy in every direction, providing a set of sightlines
less affected by the relatively local nature of some previous studies.

In the next section I describe the sample, data reduction and analysis
techniques.  In section~\ref{sec:results}, I present the results of the
study.  Section~\ref{sec:discussion} contains an analysis of the relevance
of the results and a comparison to previous efforts inside and outside our
Galaxy.  In section~\ref{sec:conclusions}, I offer my conclusions.  All errors
quoted are statistical uncertainties at the 68\% confidence level for one
parameter of interest unless stated otherwise.

\section{Sample selection, data reduction and analysis}

The aim of the work is to obtain equivalent hydrogen column densities (in
essence the metal column density, $N_{\rm H_{\rm X}}$) for a large number of
sightlines through the Galaxy from the soft X-ray photoelectric absorption
of GRB afterglows. These column densities will be
a combination of the whole column density along the line of sight to the
GRB, dominated by the Galactic column and the absorption from the GRB's host
galaxy.  This would yield a distribution of column densities with a lower
limit at the values of the Galactic column density.  By obtaining the
Galactic dust and \ion{H}{i} column densities ($E(B-V)$) and $N_{\rm HI}$
respectively) along the line-of-sight to the GRB from surveys, specifically
\citet{1998ApJ...500..525S} for Galactic dust and
\citet{2005A&A...440..775K} for \ion{H}{i}, the dust-to-metals ratio and
metallicity can be obtained by suitably fitting the two-dimensional
distributions of $N_{\rm H_{\rm X}}$--$A_V$ and $N_{\rm H_{\rm X}}$--$N_{\rm
HI}$ with a cut-off at the value of the Galactic relation. An
$R_V=3.1$ was assumed to convert $E(B-V)$ to $A_V$, consistent with the mean
and median $R_V$ values found by \citet{2007ApJ...663..320F} for 328 sources
to distances up to $\sim5$\,kpc distributed across the Galactic plane.

The dust column values were obtained from \citet{1998ApJ...500..525S} and
reduced by 14\% following the analysis of \citet{2010ApJ...725.1175S}, with
uncertainties based on the standard deviation of nearby points.  The atomic
hydrogen column densities were obtained from the Leiden-Argentine-Bonn
survey reported in \citet{2005A&A...440..775K} and accessed through the
\texttt{nH} tool in Ftools.  Uncertainties in the \ion{H}{i} column density
were set at 10\% \citep{2010arXiv1012.5319W}. 

To derive reddenings, \citet{1998ApJ...500..525S} use the $100\,\mu$m sky emission maps from
COBE/DIRBE and IRAS/ISSA with temperature corrections.  The conversion from
far-infrared (FIR) flux to reddening is made using the excess colours of a
sample of elliptical galaxies.  The zodiacal light contribution is removed
using the DIRBE $25\,\mu$m maps.  The values of E($B-V$) reported should be
fairly accurate, since the conversion from FIR emission is measured on
reddenings, with systematic uncertainties at a level below our ultimate
statistical uncertainty \citep{2010ApJ...725.1175S}.  The accuracy of using
a conversion from reddenings to absolute extinction in the Galaxy is a
matter of significant debate \citep[e.g.][]{2007ApJ...663..320F}, and will
evidently be different along different lines of sight.  However, the
standard deviation found in the conversion, i.e.  in $R_V$, is 0.27
\citep{2007ApJ...663..320F}.

The 21\,cm maps \citep{2005A&A...440..775K} are the combined Leiden/Dwingeloo
\citep{1997agnh.book.....H} and Instituto Argentino de Radioastronomía
\citep{2005A&A...440..767B} surveys detecting \ion{H}{i} over a velocity
range of $-450$\,km\,s$^{-1}$ to $+400$\,km\,s$^{-1}$ at 1.3\,km\,s$^{-1}$
resolution with very high equivalent main beam efficiency ($\gtrsim0.99$).

To obtain an equivalent hydrogen column density, $N_{\rm H_X}$, from
low-resolution X-ray spectroscopy, the deviation from a power-law at low
energies is measured.  This deviation is due to photoelectric absorption by
metals (primarily O, C, Si, Fe and He depending on the energy), and is fit
with absorption models of the gas.  I assume a neutral gas, although the
total column density is not strongly affected by the first few ionisations
of the metals since the absorption is due to inner shell electrons. 
Measured cross-sections are used for the elements, and while improvements to
the cross-sections over the years have improved fits to absorption data
\citep[][WAM00]{2000ApJ...542..914W}, the effects on low-resolution
spectroscopy are not large, a few percent at most on the total column
density (WAM00).  The most important effect is the assumed abundance of the
elements relative to hydrogen and this is discussed in
section~\ref{sec:results} below.  The relative abundances of the various
metals also has some effect on the total derived column density, however, at
these resolutions, there is no way to distinguish which elements dominate
the absorption, and since the majority of these elements contribute to the
dust, it is not a crucial point.

Recent work suggests that the hot intergalactic medium (IGM) might
increase the X-ray absorption smoothly with redshift at a low level
\citep{2011ApJ...734...26B}.  However 1) any such effect is small, 2) does not
appear to be consistent with IGM absorption since no excess absorption is detected
in about a third of quasars examined as a comparison sample, and 3) may be
related to inadequately modelled Galactic absorption.  Apart from smooth IGM
absorption, intervening galaxies are unlikely to contribute much to the
X-ray absorption in the general case as the observed absorption is related
to the total metal column density and drops approximately as $(1+z)^{2.5}$
due to bandpass effects.  Every line of sight would have to have the
equivalent of a $0.1Z_{\sun}$ metallicity absorber at redshift $z=1$ with a
neutral gas column density of $\sim5\times10^{21}$\,cm$^{-2}$ in order to
begin to impact the result here.  Such large column density intervening systems
do not exist along most sightlines, as argued in \citet{2007ApJ...660L.101W}.

All GRBs observed with \emph{Swift}'s X-ray telescope (XRT) were included in
the sample up to the end of November 2010 (GRB\,101030A), resulting in 638
GRBs.  While X-rays are detected for almost all GRBs observed by
\emph{Swift}-XRT, a significant fraction do not have sufficient
signal-to-noise to provide a spectrum good enough to determine the X-ray
absorption.

The method adopted here has been to take all pre-reduced spectra from the
Swift/XRT GRB spectrum repository \citep{2009MNRAS.397.1177E} for both the
windowed timing (WT) and photon counting (PC) modes using the appropriate
corresponding calibration files from that archive. These data were then
fit with a power-law with photo-electric absorption (\texttt{phabs(pow)} in
Xspec). 

While the PC and WT mode spectra for a given GRB are not independent
objects, they are independent measurements of the same sightline.  It is
well known that the absorption in the X-ray afterglows of some GRBs appears
to decrease as a function of time
\citep[e.g.][]{2005A&A...442L..21S,2007A&A...462..565G,2007ApJ...654L..17C}. 
For this reason, it is often preferable to choose the later PC data, as it
is likely to be closer to the Galactic value.  However, the WT data often
has considerably higher signal.  Therefore, where the PC and WT mode data
gave results consistent within $1\sigma$ (68\% confidence), both values were
used.  Where the results were discrepant at $>1\sigma$, the PC value was
used.  Since the adopted method depends solely on the limit of the
population in the $N_{\rm H_{\rm X}}$--$A_V$ plane, the result is,
however, not very sensitive to this.

The cut-off, which is the relation between the X-ray column density and the
\ion{H}{i} or $A_V$ columns, was obtained from the data using a
two-dimensional two-sample KS (2D2SKS) test.  The 2D distribution was fit
using a log-normal for the Galactic column density distributions (either
$A_V$ or $N_{\rm HI}$) and the sum of the normalised Galactic column and an
extragalactic column for the GRB afterglows, where a log-normal was also
used to reproduce the extragalactic column density distribution.  The
parameters for the Galactic column density log-normal distribution was
obtained by fitting the $A_V$ or $N_{\rm HI}$ distributions directly.  The
parameters of the extragalactic log-normal distribution as well as the
normalisation of the Galactic component of the GRB afterglow absorption
distribution were left as free parameters in the 2D2SKS fit.  This
normalisation parameter is the ratio between the X-ray absorbing column
density and the $A_V$ or $N_{\rm HI}$.  The fit was obtained by applying the
2D2SKS test at each step while stepping through each parameter across the
search space, until a maximum in the probability was obtained. 
Uncertainties in the parameters were obtained using a Monte Carlo technique
based on a Gaussian distribution of the errors on each datapoint.  The same
method was employed to obtain the best-fit $N_{\rm HI}$--$N_{\rm H_{\rm X}}$
and $A_V$--$N_{\rm H_{\rm X}}$ relations.

\section{Results}
\label{sec:results}

The best-fit to the $A_V$--$N_{\rm H_{\rm X}}$ relation is $N_{\rm H} =
2.2^{+0.3}_{-0.4}\times10^{21}$\,cm$^{-2}$\,$A_V$.  The best-fit for
the ratio of X-ray to 21\,cm absorbing column densities is $N_{\rm H_{\rm
X}}$/$N_{\rm HI} = 1.1^{0.2}_{-0.1}$.

The results obtained are, not surprisingly, somewhat sensitive to finding a
function to reproduce the distributions with reasonable fidelity.  However,
whether a normal or log-normal distribution is used to reproduce the
extragalactic column density distribution does not affect the results
significantly.  This is because the fits are sensitive to where the limit of
the 2D dataset is, rather than the precise shape of the distribution.  As
long as the distribution as a whole is reproduced with reasonable accuracy,
the nature of the function is not very important.

In all of these fits the solar metallicity estimate of
\citet[][AG89]{1989GeCoA..53..197A} were used.  It is now generally accepted
that this metallicity is $\sim45$\% higher than indicated by direct
measurements of the solar spectrum \citep{2009ARA&A..47..481A} and possibly
the local ISM (WAM00).  It was suggested by WAM00 that X-ray absorption
studies should use among other things, updated absorption cross-sections
and, in particular, updated ISM metallicity values to determine gas column
densities in the Galaxy.  The other improvements suggested by WAM00 have a
relatively small effect on the determined gas column densities, compared to
the change in metallicity they propose.  Their proposed ISM abundances
typically increase the equivalent hydrogen column density for a given
observation by roughly the inverse of the change in the metallicity.  We
fitted all of our data again using the proposed absorption model of WAM00,
i.e.\ \texttt{tbabs}, with abundances set at the levels suggested in that
paper for the ISM.  As expected, we obtain values of the $N_{\rm
H_{\rm X}}/A_V$ and $N_{\rm H_{\rm X}}$/$N_{\rm HI}$ ratios that are
$\sim40\%$ higher than using AG89 metallicities with similar fractional
uncertainties.  However, previous works have assumed metallicities similar
to the AG89 values.  Therefore it is the results using AG89 metallicities
that we will use to compare with previous measurements.  Indeed, as
discussed further below, it appears that the metallicity of a typical
Galactic sightline is not consistent with the values proposed in WAM00.

\begin{figure}
 \includegraphics[viewport=1.15 17.95 396.857941 410.8,width=\columnwidth,clip=]{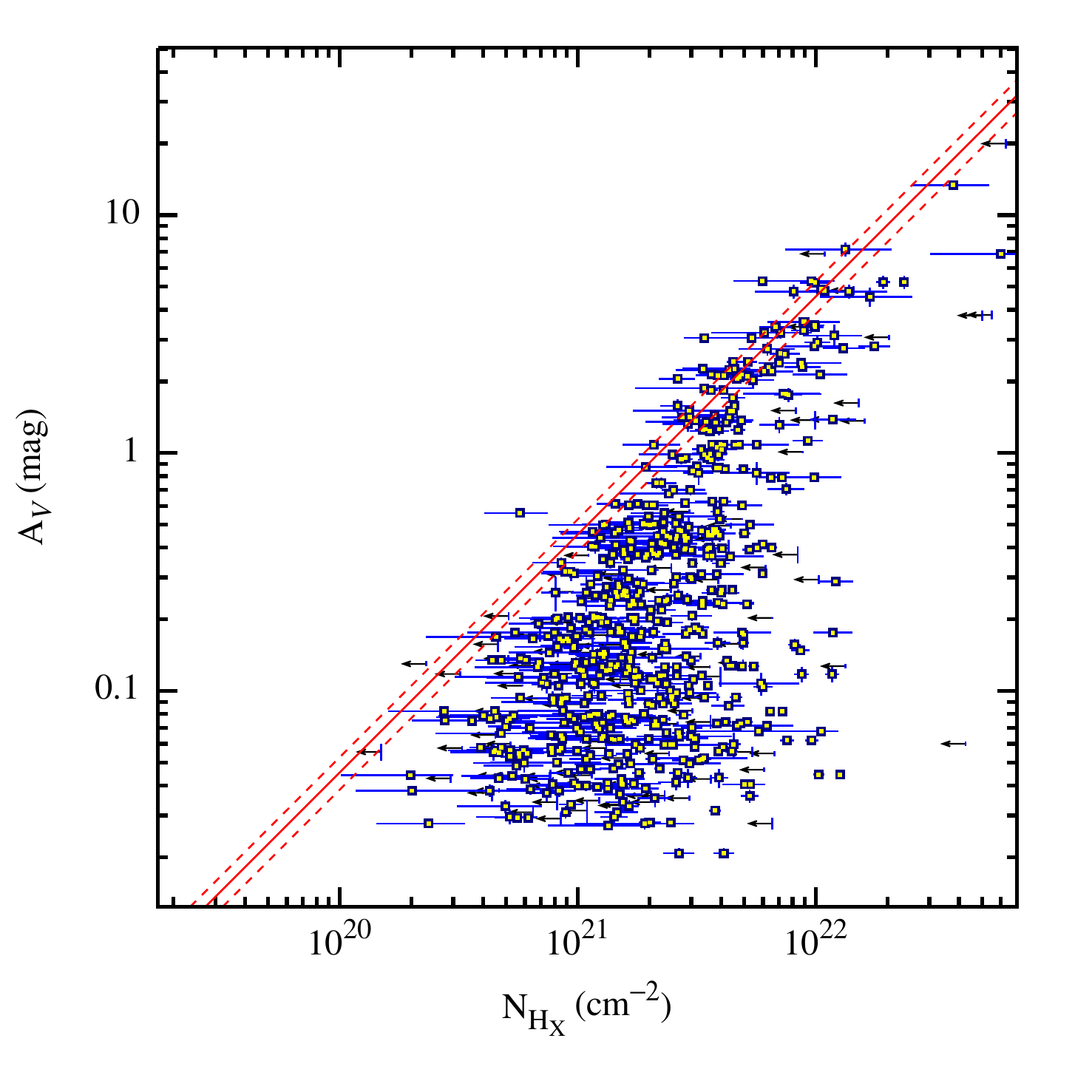}
 \caption{The Galactic dust column ($A_V$) plotted against the total equivalent hydrogen
          column density measured from soft X-ray absorption toward \emph{Swift} GRBs
          ($N_{\rm H_X}$). The best-fit limit to the population is plotted
          as a solid line with 1$\sigma$ uncertainties as dashed lines and
          represents the Galactic relation between metals and dust: $N_{\rm
          H_{\rm X}}/A_V =
          2.2^{+0.4}_{-0.3}\times10^{21}$\,cm$^{-2}$\,mag$^{-1}$, assuming
          AG89 abundances to convert the metal column to an equivalent
          hydrogen column density.}
 \label{fig:nxebv}
\end{figure}

\begin{figure}
 \includegraphics[viewport=0.15 10.95 400.857941 412.8,width=\columnwidth,clip=]{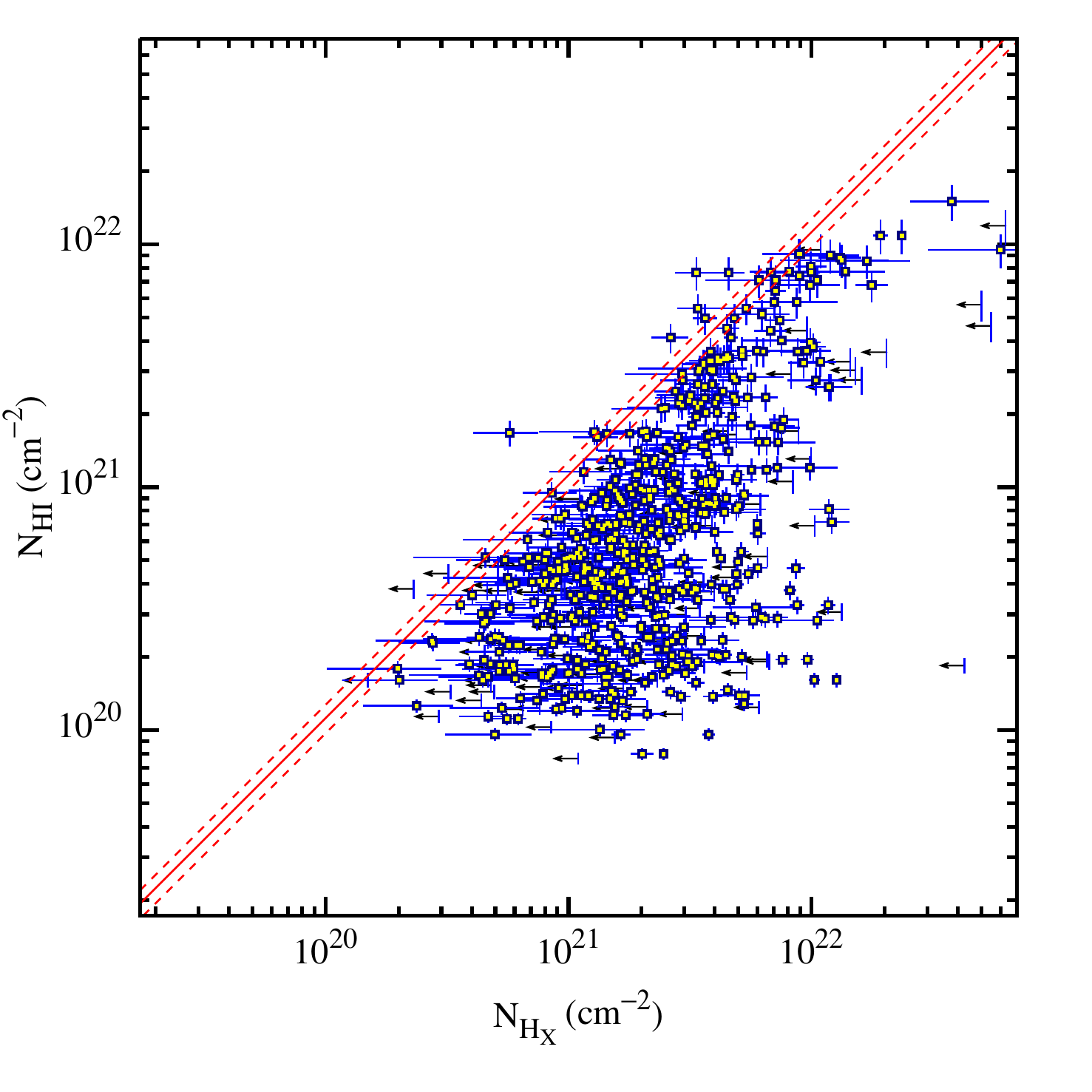}
 \caption{The Galactic \ion{H}{i} column density plotted against the total
          equivalent hydrogen column density measured from soft X-ray
          absorption toward \emph{Swift} GRBs ($N_{\rm H_X}$). While the
          soft X-ray absorption is presented in units of equivalent hydrogen
          column density, it is effectively a metal column density
          converted to equivalent column density assuming AG89
          abundances (see text).  The best-fit relation is super-solar and
          is plotted as in Fig.~\ref{fig:nxebv} above.  The best-fit relation,
          corresponding to a ratio of $N_{\rm H_X}/N_{\rm HI} =
          1.1^{+0.2}_{-0.1}$, would be even higher using the current
          \citet{2009ARA&A..47..481A}
          solar metallicities.}
 \label{fig:nxnh}
\end{figure}

\section{Discussion}
\label{sec:discussion}

The results presented here are consistent with the range of previous results
within uncertainties.  Early work focussing on bright X-ray sources,
comparing soft--X-ray metal column densities to extinction columns by
\citet{1973A&A....26..257R}, \citet{1975ApJ...198...95G}, and
\citet{1975ApJ...198..103R} provided $N_{\rm H_{\rm X}}/A_V = 1.9$, 2.2, and
$2.2\times10^{21}$\,cm$^{-2}$\,mag$^{-1}$.  Later,
\citet{1995A&A...293..889P} analysed 25 point sources and four supernova
remnants (SNRs) with data from ROSAT, deriving the soft X-ray absorption
from the spectra of the sources.  That work resulted in a value of $N_{\rm
H_{\rm X}}/A_V = 1.79\pm0.03\times10^{21}$\,cm$^{-2}$\,mag$^{-1}$.  To
extend the sample to higher extinction lines of sight,
\citet{2003A&A...408..581V} analysed the X-ray absorption and $J$-band
extinction toward six nearby star-forming regions (dominated by observations
of $\rho$\,Oph) using pre-main sequence stars.  They obtain a relation
$N_{\rm H_{\rm X}}/A_J = 5.6\pm0.4\times10^{21}$\,cm$^{-2}$\,mag$^{-1}$. 
Converting this to $N_{\rm H_{\rm X}}/A_V$ using the mean Galactic
extinction curve \citep[e.g.][]{1989ApJ...345..245C} with an $R_V=3.1$,
yields an extremely low value, $N_{\rm H_{\rm X}}/A_V \sim
1\times10^{21}$\,cm$^{-2}$\,mag$^{-1}$.  In that paper the unusually low
metals-to-dust ratio is ascribed to either low metallicity in the region of
the Galaxy local to the Sun, or to a very flat extinction curve in the
$\rho$\,Oph cloud.  The former hypothesis seems implausible since it
requires that the gas-to-dust ratio remains constant while the
metals-to-dust ratios and metallicity are lower by a factor of two.  It
therefore seems considerably more plausible that the dust grains in a dense
molecular cloud are larger, and that therefore the extinction curve is
simply flatter.  Adopting $R_V=4$ \citep{1993AJ....105.1010V} results in
$N_{\rm H_{\rm X}}/A_V = 1.7\pm0.1\times10^{21}$\,cm$^{-2}$\,mag$^{-1}$;
adopting $R_V=6$ \citep{2003A&A...408..581V} gives $1.9\times10^{21}$ with a
similar error.  Observations of 38 sightlines in the far-UV allowed
\citet{2009ApJS..180..125R} to estimate the gas-to-dust ratio including the
effects of molecularisation of the hydrogen.  They found that for their
sightlines, with E($B-V$) in the range 0.17--1.08, the fraction of hydrogen
in molecular form, $f_{H_2}$, was typically $\sim 0.5$ with none approaching
$f_{H_2}\sim1$.  Their total gas-to-dust ratio was consistent with previous
measurements, $N_{\rm H_{\rm X}}/A_V =
2.15\pm0.14\times10^{21}$\,cm$^{-2}$\,mag$^{-1}$. Finally, a recent paper
by \citet{2009MNRAS.400.2050G} analysed this issue very systematically by
selecting 22 Galactic SNRs, deriving the metal absorption from the soft
X-ray spectra and the extinction from the Balmer decrement in most cases. 
They obtained $N_{\rm H_{\rm X}}/A_V =
2.21\pm0.09\times10^{21}$\,cm$^{-2}$\,mag$^{-1}$.  In that work they suggest
that while 143 SNRs exist in the \emph{Chandra}, XMM-\emph{Newton}, and
\emph{Suzaku} archives that permit good measurements of $N_{\rm H_{\rm X}}$,
only the 22 they use have reasonable dust column density measurements
available.  An obvious extension of that work would be to obtain extinction
measurements for the remaining SNRs and increase the sample size from
$\sim20$ to over a hundred. A potential improvement of that
technique could be to do more than use the H$\alpha$/H$\beta$ Balmer
decrement which gives a measure only of the optical reddening.  If the
technique could be extended to other line ratios \citep[as recently done for
GRB host galaxies,][]{2011MNRAS.414.2793W} it would significantly improve
the understanding of the extinction properties of the dust columns along the
line of sight.

The method used in this paper possesses several advantages over previous
studies.  First, the underlying X-ray source spectrum is well-defined, almost invariably
dominated by a power-law.  Second, the sight-lines are passed at random
through the Galaxy.  Third, the sight-lines are to objects outside the
Galaxy, allowing us to include the entire Galactic column in any given
direction, and is therefore a good census of the whole Galaxy.  However, the
drawback is clearly that we are dealing with an upper-limit and therefore a
large fraction of our more than 600 sightlines do not contribute much
statistical power to the constraint on the metals-to-dust ratio.  Partly for
this reason, the uncertainties on the best-fit linear relation are fairly
large, $\sim 20\%$.

A curious fact among previous measurements deriving the
metals-to-extinction ratios from X-ray data is that the quoted statistical
errors are typically a few percent, while clearly the statistical quality of
the fits are low, with very large outliers in terms of contributions to the
fit statistic \citep[e.g.\ SNR G0.0+0.0 in][shows a very large deviation
from the fit]{2009MNRAS.400.2050G}.  From this fact alone it is obvious that
either the statistical or systematic uncertainties in these studies are
substantially underestimated or that the intrinsic variation in the
metals-to-extinction ratios is larger than the uncertainty.  In addition, the
studies quoted above reach mean values that differ by significantly more
than a few percent, e.g.\ the studies of \citet{2009MNRAS.400.2050G} and
\citet{1995A&A...293..889P} differ by more than $4\sigma$, supporting the
fact that the quoted statistical uncertainties do not represent the actual
scatter in the data.  The study on the $\rho$\,Oph cloud is instructive in
regard to deciding the origin of the scatter \citep{2003A&A...408..581V},
providing what appears to be a good quality fit for a single region in the
Galaxy, and hints that perhaps it is the inherent scatter in the
metals-to-extinction ratio that is responsible for the large deviation from one
study to the next.

\subsection{The effect of metallicity}

The results presented here are compatible with previous observations since
as far as can be determined, previous X-ray measurements used metallicity
values similar to AG89 to determine the equivalent hydrogen column density. 
The comparison made in this paper with the 21\,cm \ion{H}{i} measurements
indicates that the mean ISM metallicity in the Galaxy is approximately the
AG89 value, rather than the current best estimate of the solar metallicity
as proposed by WAM00.  While the uncertainties presented in this study are
relatively large, a more telling comparison is with Ly$\alpha$ studies of
gas-to-dust ratios in the Galaxy
\citep[e.g.]{1978ApJ...224..132B,1981MNRAS.196..469W,1994ApJ...427..274D},
which suggest a value of $N_\ion{H}{i}$ in the range
$1.6-1.9\times10^{21}A_V$\,cm$^{-2}$.  The ratio of the typical
X-ray--derived value to the Ly$\alpha$-derived value is around 1.2,
immediately indicating that the mean Galactic ISM metallicity is somewhat
\emph{larger} (approximately 20\%) than the solar metallicity values of
AG89.  Such a result is a little surprising given the debate in the X-ray
literature on the correct dust-to-gas ratio.  However, it has been known for
many years that the ISM metallicity increases toward the Galactic centre and
that the ISM metallicity becomes similar to the AG89 solar value at
around 6\,kpc from the Galactic centre \citep{2000A&A...363..537R}.  Since
the ratios derived in X-rays are dominated by high-extinction sightlines, it
is perhaps after all not that surprising that the metallicity obtained for a typical
sightline through the Galaxy is $\sim75\%$, or 0.25\,dex, higher than the
most recent measurements of the solar metallicity
\citep{2009ARA&A..47..481A}.  This value is particularly useful since it is
a direct measure of the ISM itself without having to use stellar
photospheres and it is effectively independent of dust depletion or
ionisation in the ISM.

\subsection{Proxies for the soft X-ray absorbing column density}

For a typical X-ray sightline through the Galaxy, therefore, a more accurate
estimate of the soft X-ray absorption may be obtained by using a metallicity
value somewhat higher than that of AG89.  Certainly it seems that using the
current solar metallicities \citep{2009ARA&A..47..481A} with \ion{H}{i}
column densities is likely to lead to a large underestimate of the
Galactic soft X-ray absorption.

\begin{figure}
 \includegraphics[viewport=1.152000 29 396.857941 378,width=\columnwidth,clip=]{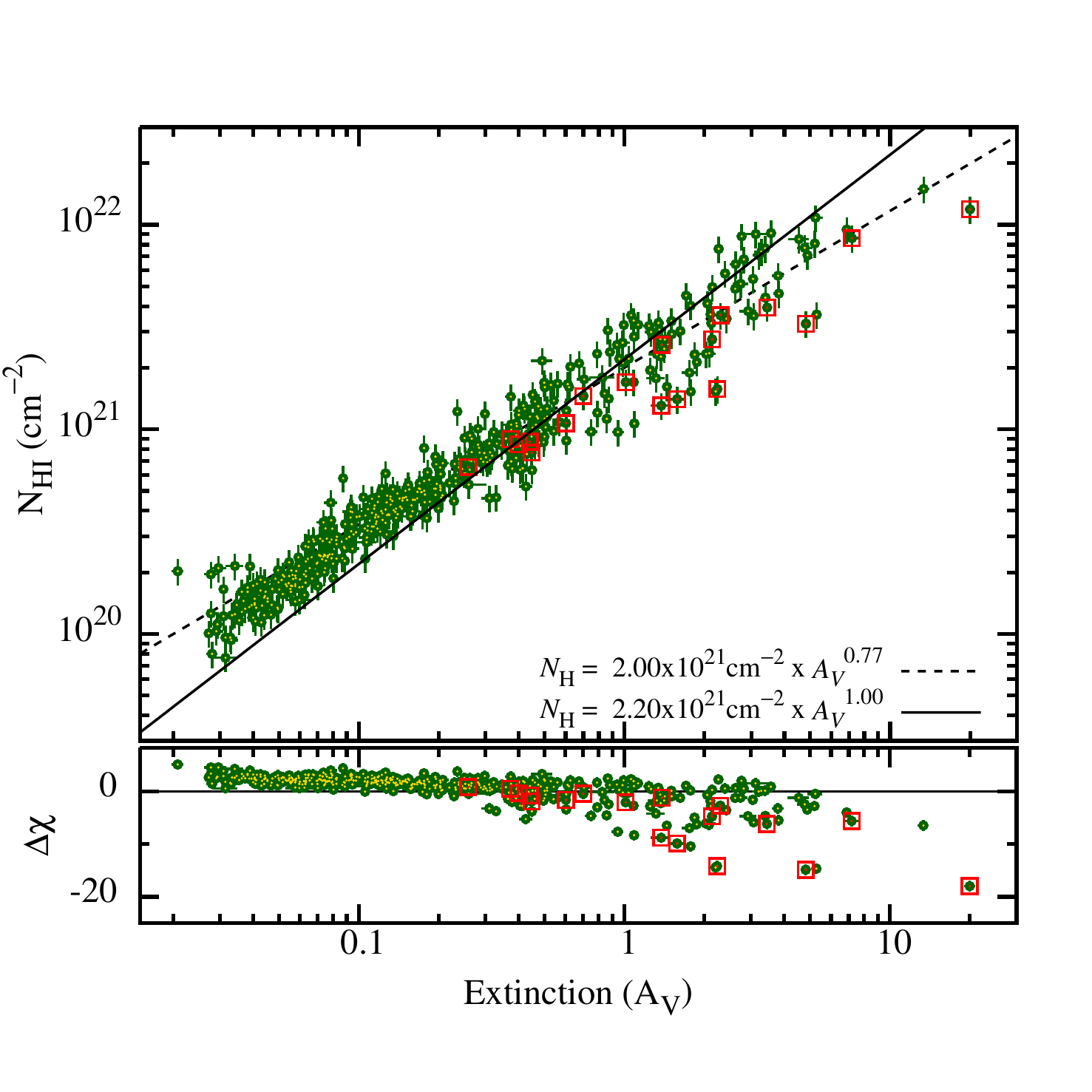}
 \caption{Atomic hydrogen column density plotted as a function of extinction
          for the sightlines associated with every GRB in the sample from
          the LAB \ion{H}{i} survey \citep{2005A&A...440..775K} and the dust
          maps of \citet{1998ApJ...500..525S}.  The anticipated linear
          relation is plotted as a solid line with residuals to this
          relation in terms of $\Delta\chi$ in the lower panel.  The linear
          relation is clearly a poor fit to the data, suggesting that the
          dust-to-gas ratio increases as a function of column density.  The
          best-fit non-linear relation is plotted as a dashed line.  The
          non-linearity in the relation may be due to metallicity variations
          along different sightlines.  The metallicity gradient in the
          Galaxy suggests that sightlines toward the Galactic centre should
          be more metal rich and therefore show a higher dust-to-gas ratio. 
          Sightlines toward the Galactic centre ($\cos(l)\cos(b)>0.9$) are
          plotted with open squares and do indeed show a higher dust-to-gas
          ratio than sightlines with comparable column densities.  }
 \label{fig:nhebv}
\end{figure}

Since the Galaxy is known to show a strong metallicity gradient
\citep{2000A&A...363..537R}, if the dust-to-metals ratio is roughly constant
\citep{2003ARA&A..41..241D}, one would expect the dust-to-gas ratio to vary
in a similar way, i.e.\ a higher dust-to-gas ratio toward the Galactic
centre.  Sightlines toward the Galactic centre will have higher column
densities: we therefore may expect a relationship in the Galaxy between the
dust-to-gas ratio and the total column density.  Such a relationship is
possibly indicated by \citet{1998ApJ...500..525S}, where the Galactic plane
has a significantly higher dust-to-gas ratio than at high-latitudes with the
Galactic centre direction being especially high.  In Fig.~\ref{fig:nhebv}
the relationship between the Galactic 21\,cm \ion{H}{i} and dust column
densities is shown.  Sightlines toward the Galactic centre
($\cos(l)\cos(b)>0.9$, open squares on Fig.~\ref{fig:nhebv}) do indeed seem
to show a higher dust-to-gas ratio than sightlines with comparable column
densities along other sightlines.  The comparison of dust to gas directly
yields a non-linear relation, with higher column density sightlines showing
a higher dust-to-gas ratio.  The linear relation $N_\ion{H}{i} =
2.2\times10^{21}A_V$ shows clear residuals.  However, $N_\ion{H}{i} =
2.00\pm0.14\times10^{21} A_V^{0.77\pm0.07}$ provides a better fit to the
data, though the fit is still not good.

Previous studies have shown a similar effect. \citet{1998ApJ...500..525S}
note that the dust content of low density high velocity \ion{H}{i} clouds is
lower than in other regions of the Galaxy.  The survey of \ion{H}{i} column
densities and extinction in the UV using Ly$\alpha$ absorption and reddening
of stars of \citet{1994ApJ...427..274D} show the same effect very clearly
(their Fig.~4a and to a lesser extent 4b).

In principle, a change in the absolute to selective extinction ratio, $R_V$,
could be responsible for the apparent change in gas-to-dust ratio derived
from reddening.  However, since we observe the same effect in
Fig.~\ref{fig:nhebv} where the extinction data are derived from the dust
emission in the far-infrared (FIR), not from reddening, this seems unlikely. 
Another possibility is that dust temperatures, which strongly affect the
emitted flux in the FIR, might produce such an effect -- cooler lines of
sight would yield a spuriously lower dust column if the assumed temperature
was too high.  However, \citet{1998ApJ...500..525S} accounted for the dust
temperature in their analysis, and more compellingly, we observe the same
effect in extinction in the UV data of \citet{1994ApJ...427..274D}.

Other explanations for the correlation of gas-to-dust ratio with column
density, such as ionisation of the hydrogen at low column densities or
molecularisation at high column densities, seems unlikely since the
gas-to-metals ratio remains effectively constant across the range we study
using \ion{H}{i} observations (Fig.~\ref{fig:nxnh}). It is worth
noting, however, that the deviation observed at the high column end of
Fig~\ref{fig:nhebv} is consistent with the factor of two decrease in the
\ion{H}{i} column density observed in far-UV molecularisation studies, where
a approximately half of the hydrogen exists in molecular form along
sightlines with moderate or greater extinctions
\citep{1978ApJ...224..132B,2009ApJS..180..125R}. Even if molecularisation
or ionisation were partly responsible, however, it does not alter the fact
that the \ion{H}{i} column density is not the best proxy for the soft X-ray
absorbing column density.

This result immediately suggests that while using $N_\ion{H}{i}$ as a proxy
for the soft X-ray absorption is reasonable, using the dust column, as
suggested by \citet{1998ApJ...500..525S}, is likely to yield better results not simply
because of the higher resolution of the dust maps, but because the
dust-to-metals ratio seems to be more constant than the dust-to-gas ratio.

While this cannot be investigated further here due to the large uncertainty
in the ratios we derive, future studies should try to determine the
systematic Galactic radial and column density dependences of the
dust-to-metals ratio.  This latter point may be important since it is
well-known that the depletion of metals out of the gas phase increases as we
move into the disk and into cooler environments \citep{1996ARA&A..34..279S},
which would obviously be associated with higher column densities. 
It has been suggested above that the exploration of
\citet{2009MNRAS.400.2050G} could
be extended by increasing the sample size.  Such an extension would indeed
be worthwhile and allow the Galaxy to be divided into various lines of
sight, related to the disk or the bulge.  A more complete approach might be
to use bright extragalactic objects known to have low host galaxy
absorptions, which would allow a census to be taken of sightlines through
arbitrary directions in the galaxy, including the halo.  An obvious
candidate are blazars, where the dust extinction could be derived either as
done here from Galactic dust maps, or from reddening of the optical
power-law itself (where the caveats are that the data in different bands be
taken simultaneously and that one needs to account for intrinsic curvature
of the spectra).  Indeed, for a smaller sample, the X-ray and some of the
optical-UV data already exists.  Blazars might be bright enough for UV
spectroscopy as well, allowing H$_2$ and Ly$\alpha$ measurements to be made. 
Such a programme, while observationally intensive for a large enough sample
to probe a significant fraction of the Galaxy, could represent one of the
best probes of the Galactic ISM yet devised.

\subsection{Extragalactic metals-to-dust ratios}
Studies of metals-to-dust and gas-to-dust also exist outside the local
group.  Metals-to-dust ratios from foreground lensing galaxies at
$z\lesssim1$ have been obtained using multiply-imaged quasars
\citep{2006ApJ...637...53D, 2009ApJ...692..677D}.  Results show
metals-to-dust ratios consistent with Galactic values. Those objects are typically relatively high mass, evolved
systems.  \citet{2011arXiv1102.1469Z} showed in an analysis of GRB afterglows that GRB
host galaxies, which are typically young and star-forming with low
metallicities and hard radiation environments
\citep{2004A&A...425..913C,2006ApJ...653L..85C,2009ApJ...691..182S,2010MNRAS.405...57S,2002ApJ...566..229C,2008ApJ...683..321F,2010arXiv1010.1783W}, have substantially
lower dust-to-gas ratios than the local group even after accounting for
metallicity.

\section{Conclusions}
\label{sec:conclusions}

An analysis of the dust-to-metals ratio and metallicity of sightlines
through the Galaxy has been presented.  The Galactic metal column densities
were determined using the lower bound of the distribution of soft X-ray
absorptions of the afterglows of a large sample of GRBs detected by the
\emph{Swift} satellite.  The corresponding extinction and gas column
densities were found using the dust and \ion{H}{i} maps of
\citet{1998ApJ...500..525S} and \citet{2005A&A...440..775K} respectively. 
The metal to atomic hydrogen relation is well reproduced with a metallicity
$\sim1.75$ times the solar metallicity of \citet{2009ARA&A..47..481A}.  The
best-fitting relation between metal and dust column densities is
$N_{H_X}/A_V = 2.2_{-0.3}^{+0.4}\times10^{21}$\,cm$^{-2}$\,mag$^{-1}$ (using
AG89 abundances).  Previous observations are consistent with this result,
suggesting that the metallicity for a typical ISM sightline is 0.25\,dex
higher than the current best value for the solar metallicity.  It is
therefore suggested that a better reproduction of the Galactic soft X-ray
absorption will be provided with a metallicity $\sim20\%$ \emph{higher} than
the solar metallicity of AG89.

However, it is also found that a linear representation does not reproduce
the gas-to-dust relationship very well.  A gas-to-dust relationship with
$N_\ion{H}{i} = 2.00\pm0.14\times A_V^{0.77\pm0.07}$ provides a much better
fit to the data. It is very likely that this is predominantly a
metallicity-gradient effect, and it is therefore concluded that
while the gas-to-dust relation may be as given above, the best
proxy for the Galactic soft X-ray absorption should be given by the dust column
density with a relation of $N_{\rm H_X}/A_V =
2.2\times10^{21}$\,cm$^{-2}$\,mag$^{-1}$ or $N_{\rm H_X}/E(B-V) =
6.8\times10^{21}$\,cm$^{-2}$\,mag$^{-1}$ for an R$_V=3.1$.

\begin{acknowledgements}

The Dark Cosmology Centre is funded by the DNRF.  I would like to thank Anja
C.  Andersen, Jens Hjorth, Daniele Malesani, Johan Fynbo, and Marten van
Kerkwijk for useful discussions.  This work made use of data supplied by the
UK Swift Science Data Centre at the University of Leicester.

\end{acknowledgements}

\bibliography{apj-jour,grbs}

\end{document}